\documentclass[prl,showpacs,twocolumn,floatfix,unsortedaddress,superscriptaddress]{revtex4}
\usepackage{graphicx}

\def\bea{\begin{eqnarray}}
\def\eea{\end{eqnarray}}
\def\ben{\begin{equation}}
\def\een{\end{equation}}
\def\benu{\begin{enumerate}}
\def\enu{\end{enumerate}}

\def\n{n}

\def\sss{\scriptscriptstyle\rm}

\def\1var{(\bx_1...\bx\N)}

\def\br{{\bf r}}

\def\bx{{x}}

\def\bq{{\bf q}}

\def\x{_{\sss X}}
\def\c{_{\sss C}}
\def\s{_{\sss S}}
\def\xc{_{\sss XC}}

\def\N{_{\sss N}}

\def\GGA{^{\rm GGA}}

\def\unif{^{\rm unif}}
\def\surf{^{\rm surf}}
\def\excsurf{e\xc\surf}
\def\exsurf{e\x\surf}

\def\sph_int{ {\int d^3 r}}

\input rotate

\begin{document}
\def\z{_\zeta}
\def\GE{_{GE}}
\def\x{_x}
\def\xc{_{xc}}
\def\c{_c}
\def\X{$x~$}
\def\XC{$xc~$}
\def\C{$c~$}

\title{Restoring the 
density-gradient expansion for exchange in solids and surfaces}
\author{John P. Perdew}
\author{Adrienn Ruzsinszky}
\affiliation{Department of Physics and Quantum Theory Group, Tulane University, New Orleans, LA 70118}
\author{G\'abor I. Csonka}
\affiliation{Department of Chemistry, Budapest U. of Technology and Economics, H-1521                         
Budapest, Hungary}
\author{Oleg A. Vydrov}
\author{Gustavo E. Scuseria}
\affiliation{Department of Chemistry, Rice University, Houston, TX 77005}
\author{Lucian A. Constantin}
\affiliation{Donostia International Physics Center, E-20018, Donostia, Basque Country}
\author{Xiaolan Zhou}
\affiliation{Department of Physics and Quantum Theory Group, Tulane University, New Orleans, LA 70118}
\author{Kieron Burke}
\affiliation{Departments of Chemistry and of Physics, UC Irvine, CA 92697}
\date{\today}
\begin{abstract} 
Popular modern generalized gradient approximations (GGA's) are
biased toward the description of free-atom energies.
Restoration of the first-principles gradient
expansion for exchange over a wide range of density gradients
eliminates this bias.
We introduce PBEsol, a revised 
Perdew-Burke-Ernzerhof GGA that improves equilibrium properties of 
densely-packed solids and their surfaces.
\end{abstract}


\maketitle


Ground-state Kohn-Sham density functional theory (DFT) 
\cite{KS65}
has been hugely successful
for electronic structure calculations of solids
and molecules.  
It relies upon good approximations to the exchange-correlation (xc) energy as a 
functional of the electronic spin densities\cite{FNM03}.  Over four decades, 
increasingly 
sophisticated approximations have been developed\cite{TPSS03}.  The most commonly used 
in 
solid-state calculations today is the PBE
version\cite{PBE96} of the generalized gradient
approximation (GGA), employing both the density and its
gradient at each point in space.
Popular GGAs represent a well-tempered balance
between computational efficiency, numerical accuracy,
and reliability, but PBE also juggles the demands of
quantum chemistry and solid-state physics\cite{KPB99}.

While PBE represented a high-point of non-empirical functional
development 11 years ago, 
much has since been learned about its limitations.
PBE reduces the chronic overbinding of 
the local
spin density approximation (LSDA)\cite{KS65} but,
while LSDA
often  slightly underestimates equilibrium
lattice constants by about 1\%, PBE usually overestimates them
by about the same amount.  Other equilibrium properties, such
as bulk moduli, phonon frequencies, magnetism, and ferro-electricity,
are 
sensitive to the lattice constant, and so are also overcorrected
by PBE\cite{WC06}.   Surface energies are
too low in LSDA, but are made lower still by PBE\cite{CPT06}.

However, attempts to construct a better GGA face a Procrustean
dilemma\cite{PBE98}:
Those with an enhanced gradient dependence\cite{ZY98,HHN99} improve
atomization and total energies, but worsen bond lengths,
while more recent suggestions of a GGA for 
solids\cite{WC06,VJKS00,AM05,TLBS07,VRRK07} (like the very first 
GGA\cite{LM83}) have a reduced
gradient dependence and typically do improve lattice
parameters and/or surface energies, but have been
criticized for worsening total and atomization energies\cite{TLBS07}.
More advanced functionals have been constructed.
For example, meta-GGA's, using also the orbital kinetic-energy 
densities, provide greater accuracy over a wider range of systems
and properties\cite{TPSS03}.
But current meta-GGA's do not improve lattice constants as
dramatically as surface energies, and
meta-GGA's are not yet available in all solid-state codes.

In the present work, we explain the origin of this dilemma and show
that {\em no} GGA can do both:  Accurate atomic
exchange energies require violating the gradient expansion for slowly-varying
densities, which is valid for solids and their surfaces.  At the GGA level,
one must choose.  
A pragmatic approach to lattice properties is therefore to
use a modified functional especially for solids which, unlike
previous suggestions, recovers the
gradient expansion for exchange over a wide range of density
gradients.
This by itself yields much better
lattice constants than PBE (see Ref. \cite{EPAPS}),
and excellent jellium surface exchange energies, because such energies
are dominated by moderate density gradients.
Numerical studies\cite{FNM03} suggest that PBEsol will also yield
highly accurate bulk exchange energies of metals within the pseudopotential approximation.
These observations suggest an improved starting point for more advanced functional 
construction.

Restoration of the gradient expanson for exchange
requires a complementary alteration for correlation.
We fit the jellium surface xc energy,
as done by Armiento and Mattsson\cite{AM05}.  This condition is
insufficient to yield accurate lattice constants, as experience
with the TPSS meta-GGA shows\cite{TPSS03}.  But, as explained below,
it is relevant to lattice constants,
because of its bulk-like nature.  
Being accurate for both surface exchange and xc energies,
PBEsol minimizes reliance on error cancellation between x and c, while
Ref. \cite{AM05} does not.
By using the PBE form but
simply altering two parameters,
we retain all other exact conditions that make PBE so reliable.

The GGA form for the exchange energy is simply
\ben
E\x\GGA[\n] = \int d^3r\, e\x\unif(\n(\br))\, F\x(s(\br))
\een
where $\n(\br)$ is the electronic density, $e\x\unif(n)$ is 
the exchange energy density of a uniform electron gas (
$\sim n^{4/3}$), $s=|\nabla \n|/(2 k_F \n)$ (with $k_F=
(3\pi^2\n)^{1/3}$) is the dimensionless density gradient, and
$F\x(s)$ is the enhancement factor for the given GGA\cite{PBE96}.
Eq. (1) is the spin-unpolarized form, from which 
the spin dependence can be deduced\cite{PBE96}.
Any GGA that recovers the uniform gas limit has
\ben
F\x(s) = 1 + \mu\, s^2 + \ldots.~~~~~(s\to 0)
\een
The gradient expansion\cite{KS65}
that is accurate for slowly-varying electron gases has\cite{AK85}
$\mu\GE =10/81 \approx  .1235$.

To begin,
Ref \cite{PCSB06} showed that the
exchange energies of neutral atoms are very well approximated
by their asymptotic expansion for large $Z$, i.e., $E\x =
-.2208 Z^{5/3} -.196 Z +...$.   The first term arises from
LSDA, but the second arises in a GGA from the $s^2$
contribution to Eq. (2) and requires $\mu\approx 2 \mu\GE$.
Thus any GGA that is accurate for the exchange energies of free neutral
atoms must have $\mu\approx 2\mu\GE$.
PBE does, although its value 
of $\mu=0.2195$ was
found from a different non-empirical argument.  So does B88, as it
was fitted to the \X energies of noble gas atoms\cite{B88}.  
Even PW91 does\cite{PBW96}, as it reverts only at irrelevantly small $s$ to
$\mu\GE$.



Thus, to attain accurate exchange energies
of atoms (vital to dissociation
energies in molecules and cohesive energies in solids),
any GGA must strongly violate the gradient expansion for
slowly-varying densities\cite{PCSB06}.
But most of thermochemistry occurs without free atoms, and is not much worse
in LSDA than in PBE (e.g., \cite{Bb06}). 
Moreover, for the evaluation of exchange, the
densities of real solids and their surfaces are often
almost slowly-varying over space.  Restoring the gradient
expansion should improve their description (but 
worsen atomization energies).  The GGA is a limited form, and {\em
cannot} satisfy both conditions.  
Eq. (2) suggests a necessary condition for convergence of the second-order          
gradient expansion for exchange: $s \lesssim 1$.  Since 
$s \lesssim 1$  for valence electrons in densely-packed 
solids (or $s \lesssim 2$ in core-valence regions of alkali atoms),
and since the reduced Laplacian of the
density is also $\lesssim 1$, the gradient expansion is
important for exchange in solids.
We choose $\mu\GE$ for PBEsol.

Now, for a GGA correlation functional that recovers the uniform gas limit,
the gradient expansion is 
\ben
E\c[\n] = \int d^3r\, \n(\br)\, \{ \epsilon\c\unif(\n(\br))+\beta\, t^2(\br)\, + \ldots\}
\een
where
$ \epsilon\c\unif(\n)$ is the correlation energy per particle of the
uniform gas, $\beta$ is a coefficient, and 
$t =|\nabla n|/\{2\, k_{TF}\, \n\}$ is the appropriate
reduced density gradient for correlation
(fixed by the Thomas-Fermi screening
wavevector $k_{TF}={\sqrt{4 k_F/\pi}}$,
not $k_F$.)
For slowly-varying high densities\cite{MB68},
$\beta\GE=0.0667$.
Unlike exchange, the second-order term in      
the gradient expansion for correlation cannot be small
 compared to the local term everywhere even for valence electrons in solids:
 $\beta t^2$ can be large compared to $|\epsilon\c\unif|$
(as $\beta\GE t^2 = 0.1s^2/r\s$).  The gradient expansion
can be relevant to real systems (especially solids) for
exchange, but much less so for correlation.

Also relevant to solids is
$f\xc(q)$ for the response of the uniform gas
to 
a weak potential
$\lambda\, \cos (\bq\cdot\br)$.  For valence-electron densities,
the exact $f\xc(q)$ is
almost independent of $q$, up to
$2k_F$ \cite{MCS95}.
Thus LSDA, which produces a constant (the value at $q=0$),
yields an accurate approximation 
for $q \lesssim 2 k_F$.  But any GGA with a
non-zero \XC contribution to second order in $\nabla n$
produces a term quadratic in $q$.
Since we are interested in weakly-varying valence
electron densities in densely-packed solids, we wish to retain this
excellent feature of LSDA.  If
\ben
\mu = \pi^2 \beta/3,
\label{mubet}
\een
there is complete cancellation between beyond-LSDA \X and \C 
contributions, restoring LSDA response.

In PBE, the gradient expansion for correlation is respected, i.e.,
$\beta=\beta\GE$, and $\mu \approx 2\mu\GE$ satisfies Eq. (\ref{mubet}).
This choice agrees well
with
the PW91 exchange functional, 
and with its hole construction\cite{PBW96} (where sharp real-space cutoffs are most 
appropriate 
to atoms), and yields
highly accurate exchange energies of
atoms.
But we have already argued that $\mu \approx 2\mu\GE$ is harmful for
many condensed matter applications.
Once we choose $\mu\GE$ for exchange, we cannot 
recover simultaneously 
the GEA for correlation and the linear response of LSDA for a 
uniform density.  Exact satisfaction of Eq. (\ref{mubet}) would
yield $\beta=0.0375$, 
but a compromise value will satisfy
another, more relevant constraint for solid-state
applications.


For correlation, large neutral jellium clusters are our paradigm,
for which
$E\xc \to e\xc\unif V + \excsurf A + ..$ as the radius grows,
where $\excsurf$ is the jellium surface \XC energy, $V$ the
volume of the cluster and $A$ its area.
A GGA that recovers $\excsurf$ will be 
correct in leading- and next-order for neutral jellium
clusters as $N\to\infty$, in a similar way to 
popular exchange GGA's
for neutral atoms.
Moreover, the surface energy is
dominated by \XC contributions and $\excsurf$ is
a bulk-like property, arising mainly (103\% at $r\s=2$) from
a moderately-varying-density region (with $s \lesssim 1.9$) inside
the classical turning plane. 

We check that this condition is compatible with the restoration
of the gradient expansion for exchange.  Because jellium clusters have
a uniform bulk density and because
most of the surface energy comes from within, the
gradient expansion should be accurate.
We find, at bulk density $r\s=3$, the 
errors of the surface exchange energy are: LSDA 27\%, PBE -11\% and PBEsol 2.7\%.

\begin{figure}[htb]
\unitlength1cm
\begin{picture}(12.5,5.5)
\put(-7,9){\makebox(12,6.5){
\includegraphics{rat.ps}
}}
\end{picture}
\caption{Ratio of calculated surface exchange-correlation
to that of LSDA as a function of $r\s$
for various approximations.}
\end{figure}

We fit $\excsurf$\cite{AM05} to determine
our correlation functional.
But the jellium $\excsurf$ is not known exactly.
Figure 1 shows surface energy enhancements relative
to LSDA.
The likely ``range of the possible''         
for $\excsurf$ extends from TPSS
meta-GGA
\cite{TPSS03,CPT06} 
or RPA+
 \cite{YPK00}
at the low end of what is possible (in agreement with the most
recent Quantum Monte Carlo calculations\cite{WHFG07})
to the RPA-like Pitarke-Perdew (PP)\cite{PP03}
value at the high end.
For a history of this subject, and another sophisticated
estimate, see Ref. \cite{CPDG08}.
TPSS may provide the best target for a GGA.  We choose
$\beta=0.046$
and $\mu=\mu\GE$ (within the PBE form) for PBEsol, to best fit the TPSS results.
PBEsol should improve most
surface energies over LSDA, whereas PBE worsens them.

Thus we have violated Eq. (\ref{mubet}) in favor
of good surface energies.
But our value for $\beta$ is considerably closer to that
of the linear response requirement (0.0375) than that demanded by
complete restoration
of the gradient expansion (0.0667).  

\begin{figure}[htb]
\unitlength1cm
\begin{picture}(12.5,11)
\put(-7,6){\makebox(12,11){
\includegraphics{Fxc.ps}
}}
\end{picture}
\caption{Enhancement factors of PBE and PBEsol, for
spin-unpolarized systems, as a function of reduced density
gradient, for various values of $r\s$.}
\end{figure}
PBEsol becomes
exact for solids under intense compression, where
real solids and their surfaces become truly 
slowly-varying, and exchange dominates over correlation\cite{PCSB06}.
In Fig. 2, we plot the enhancement factors of PBE and PBEsol.
For a spin-unpolarized ($\zeta=0$) density $\n = 3/(4\pi r\s^3)$, we define
$F\xc(r\s,s)$
by
\ben
E\xc\GGA[\n] = \int d^3r\, e\x\unif(\n(\br))\, F\xc(r\s(\br),s(\br)).
\een
The high-density ($r\s\to 0$) limit is $F\x(s)$ of Eq. (1).
The nonlocality or
$s$-dependence of GGA exchange is diminished from
PBE to PBEsol, making the latter
somewhat closer to LSDA.
Over the whole range $s\lesssim 1$, the PBEsol
$F\x$ is close to $1+\mu\GE s^2$.
  The range
$0 \lesssim s \lesssim 3$
is energetically important for
most properties of most real systems, while
$0\lesssim s\lesssim 1$
and $1< r\s <10$ are the ranges for
valence-electron regions in many densely-packed solids.

\begin{table}[htb]
\begin{tabular}{|c|cccc|}
\hline
class&LSDA&PBE&TPSS&PBEsol\\
\hline
\multicolumn{5}{|c|}{mean error}\\
4 simple metals &-9.0&2.9&5.3&-0.3\\
5 semiconductors &-1.1&7.9&6.2&3.0\\
5 ionic solids &-8.4&8.5&6.8&2.0\\
4 transition metals &-4.0&6.4&2.5&0.0\\
total &-5.5&6.6&5.4&1.3\\
\hline
\multicolumn{5}{|c|}{mean absolute error}\\
4 simple metals &9.0&3.4&5.3&2.3\\
5 semiconductors &1.3&7.9&6.2&3.0\\
5 ionic solids &8.4&8.5&6.8&2.7\\
4 transition metals &4.0&6.4&2.7&1.9\\
total &5.6&6.7&5.4&2.5\\
\hline
\end{tabular}
\label{t:G2}
\caption{Errors in equilibrium lattice constants (in \AA$\times 10^{-2}$)
on our data set of 18 solids, relative to experiment
with estimates of the zero-point anharmonic 
expansion removed\cite{SSTP04}.}
\end{table}
To test our functional, we employ a test set of 18 solids from
Ref \cite{SSTP04}.  
These come in four groups:  Simple metals (Li,Na,K,Al),
semiconductors (C,Si,SiC,Ge,GaAs), ionic solids (NaF,NaCl,LiCl,LiF,MgO),
and transition metals (Cu,Rh,Pd,Ag).  The set is not claimed to be 
representative, but was chosen for the availability of basis functions
and anharmonic corrections\cite{SSTP04}.
Our calculations use the 
Gaussian orbital periodic code of Ref \cite{SSTP04},
with basis sets of the same or higher 
quality.
In Table I, 
we list both mean and mean absolute errors for lattice constants. The systematic PBE 
overestimate
is close to the systematic LSDA underestimate, as shown by the
total mean absolute errors, and TPSS cures this very little.
On the other hand, PBEsol greatly reduces
this overestimate, by a factor of almost 4, except for
semiconductors, where LSDA is unsurpassed.

\begin{table}[htb]
\begin{tabular}{|c|cccc|}
\hline
error&LSDA&PBE&TPSS&PBEsol\\
\hline
mean error &3.35&0.54&0.18&1.56\\
mean abs. error &3.35&0.67&0.26&1.56\\
\hline
\end{tabular}
\label{t:AE6}
\caption{Errors in atomization energies (eV) for the AE6 
set of molecules, using the 6-311+G(3df,2p) basis set.}
\end{table}
PBEsol is not expected to give good atomization energies.
In Table 2, we give the errors on the AE6 data set of molecules.
These 6 molecules (SiH$_4$, S$_2$, SiO, C$_3$H$_4$ (propyne), 
C$_2$H$_2$O$_2$ (glyoxal), and C$_4$H$_8$ (cyclobutane) were
chosen\cite{LT03}  to be representative, i.e., to reproduce the errors
of much larger data sets.  As is clear, and expected, PBEsol is much
less accurate than PBE, only about halving the error of LSDA.  
This can be related to PBEsol's worsened total energies of atoms.
 
    We have demonstrated the relevance of the second-order
 gradient coefficient for                
the exchange energy of a slowly-varying density to
the bulk and surface properties                 
of solids.  The TPSS meta-GGA \cite{TPSS03}, which incorporates this
 coefficient, gets good surface          
energies but its lattice constants are only marginally
better than those of PBE 
on which it builds, whereas PBEsol is significantly better.
This suggests that an improved meta-GGA needs to recover
the gradient expansion for exchange over a wider range of 
density distributions $n(\bf{r})$
than TPSS does.

Previous attempts to improve on PBE within the GGA form have retained the PBE
gradient coefficients $\mu$ and $\beta$
for small $s$, but altered the behavior at large
$s$\cite{ZY98,HHN99,WC06}, or have zeroed out
$\mu$\cite{VJKS00,AM05},
and are thus fundamentally different from PBEsol.
The AM05\cite{AM05} functional performs very similarly to PBEsol for the
solids studied here, but 
AM05 follows the proposal of Vitos et al.\cite{VJKS00}
to fit the conventional exchange energy density                                
of an Airy gas.
For $r\s=2$ to $6$, 
AM05 exchange has errors of 12 to 60\%
for $\exsurf$, compared to 1.6 to 4.1\% for
PBEsol.  
Numerical comparisons and details are available\cite{EPAPS}.

We have identified the simple exchange-correlation physics underlying
many properties of many solids, and shown how it differs from that for atoms.
We recommend PBEsol for the applications discussed here.
Any existing code that implements PBE can be instantly modified
to try PBEsol, by simply replacing the values of $\mu$ and $\beta$.
Modified PBE subroutines are available from http://dft.uci.edu.
We thank NSF (CHE-0355405, CHE-0457030, and DMR-0501588) and OTKA for support.

\newpage

\begin{center}
\Large
ERRATUM\\Restoring the
density-gradient expansion for exchange in solids and surfaces
\end{center}                                                        

$\;$

$\;$

   The errors of the density-functional lattice constants were 
deviations from corrected experimental values, where the corrections 
removed the finite-temperature and zero-point anharmonic expansion (ZPAE) 
effects.  The ZPAE corrections were taken  from Ref. 28 and based on 
Eq. (18) of Ref. 28, i.e., on Appendix A of A.B. 
Alchagirov et al., Phys. Rev. B $\mathbf{63}$, 224115 (2001), in which  
$\epsilon$ is the energy per atom and $v_0$ is the volume per atom.  But 
Ref. 28 inadvertently used the volume per unit cell, which is twice 
the volume per atom for the semiconductors and 
ionic crystals. Thus the ZPAE corrections for these solids in Ref. 28 
and in our article were only half what they should have been.   
Parts of Table I (with errors in units of 0.01 \AA) of our article 
should be corrected as follows:\\

%
%
\begin{tabular}{|c|cccc|}
\hline
class&LSDA&PBE&TPSS&PBEsol\\
\hline
\multicolumn{5}{|c|}{mean error}\\
5 semiconductors &-0.3&8.6&7.0&3.7\\
5 ionic solids &-6.9&10.1&8.3&3.5\\
total &-4.9&7.3&6.0&1.9\\
\hline
\multicolumn{5}{|c|}{mean absolute error}\\
5 semiconductors &0.9&8.6&7.0&3.7\\
5 ionic solids &6.9&10.1&8.3&3.5\\
total &5.1&7.4&6.0&2.9\\
\hline
\end{tabular}
%

$\;$

The corrected experimental lattice constants for the 4 simple metals 
and 4 transition metals are unaffected.  No conclusion is changed by 
this erratum, although the improvement of PBEsol relative to LSDA is 
slightly reduced.  Some lattice constants (in \AA) from the "Expt.-ZPAE" 
column of Table S1 of the Supplementary Information  (EPAPS 
PRLTAO-100-016814) should be corrected as follows: 3.544 (C), 5.416 (Si), 
4.340 (SiC), 5.640 (Ge), 5.638 (GaAs), 5.566 (NaCl), 4.579 (NaF), 
5.074 (LiCl), 3.964 (LiF), and 4.188 (MgO).  
J.P.P. thanks Fabien Tran for pointing out this error.


\begin{thebibliography}{99}

\bibitem{KS65}
W. Kohn and L.J. Sham, Phys. Rev.  {\bf 140}, A 1133 (1965).

 
\bibitem{FNM03}
U. von Barth, Physica Scripta {\bf T109}, 9 (2004).

 
\bibitem{TPSS03}
J. Tao, J.P. Perdew, V.N. Staroverov and G.E. Scuseria, Phys. Rev. Lett. {\bf 91}, 14640 (2003).



 
\bibitem{PBE96}
J.P.~Perdew, K.~Burke, and M.~Ernzerhof, Phys. Rev. Lett. {\bf 77}, 3865 
(1996); {\bf 78}, 1396 (1997) (E).
 
\bibitem{KPB99}
S. Kurth, J.P. Perdew, and
P. Blaha, Int. J. Quantum Chem.  {\bf 75}, 889 (1999).
 
\bibitem{WC06}
Z. Wu and R.E. Cohen, Phys. Rev. B {\bf 73}, 235116 (2006).

 
\bibitem{CPT06}
L.A. Constantin, J.P. Perdew, and J. Tao, Phys. Rev. B {\bf 73},
205104 (2006), and references therein.
 
\bibitem{PBE98}
J.P. Perdew, K. Burke, and
M. Ernzerhof,  Phys. Rev. Lett. {\bf 80}, 891  (1998).

 
\bibitem{ZY98}
Y. Zhang and W. Yang, Phys. Rev. Lett. {\bf 80}, 890 (1998).

 
\bibitem{HHN99}
B. Hammer, L.B. Hansen, J.K. N{\o}rskov, Phys. Rev. B {\bf 59}, 007413 (1999).
 
\bibitem{VJKS00}
L. Vitos,
B. Johansson, J. Kollar, and H.L. Skriver, Phys. Rev. B {\bf 62}, 10046
(2000).
 
\bibitem{AM05}
R. Armiento and A.E. Mattsson, Phys. Rev. B {\bf 72}, 085108 (2005).

 
\bibitem{TLBS07}
F. Tran, R. Laskowski, P. Blaha, and K. Schwarz, Phys. Rev. B {\bf 75}, 115131 (2007).

 
\bibitem{VRRK07}
F.D. Vila, J.J. Rehr, H.H. Rossner, and H.J. Krappe, Phys. Rev. B {\bf 76}
014301 (2007).
 

\bibitem{LM83}
D.C. Langreth and M.J. Mehl, Phys. Rev. B {\bf 28}, 1809 (1983).

\bibitem{EPAPS}
See EPAPS document No. XXX.  For more information on EPAPS,
see http://www.aps.org/pubservs/epaps.html.



\bibitem{AK85}
P.R. Antoniewicz and L. Kleinman, Phys. Rev. B {\bf 31}, 6779 (1985).

 
\bibitem{PCSB06}
J. P. Perdew, L. A. Constantin, E. Sagvolden, and K. Burke, Phys. Rev. Lett. {\bf 97}, 223002 (2006). 

 
\bibitem{B88}
A.D. Becke, Phys. Rev. A {\bf 38}, 3098 (1988).

 
\bibitem{PBW96}
J.P. Perdew, K. Burke, and Y. Wang,
Phys. Rev. B {\bf 54}, 16533 (1996);   {\bf 57}, 14999 (1998) (E).
 
\bibitem{Bb06}
G.I. Csonka A. Ruzsinszky, J. Tao, and J.P. Perdew, Int. J. Quantum Chem. 
101, 506 (2005).


 
\bibitem{MB68}
S.-K. Ma and K.A. Brueckner, Phys. Rev.  {\bf 165}, 18 (1968).

 
\bibitem{MCS95}
S. Moroni, D.M. Ceperley, and G. Senatore, Phys. Rev. Lett. {\bf 75}, 689 
(1995).

 
\bibitem{YPK00}
Z.D. Yan, J.P. Perdew, and S. Kurth, Phys. Rev. B {\bf 61}, 16430 (2000).
 
\bibitem{WHFG07}
B. Wood, N.D.M. Hine, W.M.C. Foulkes, and P. Garcia-Gonzalez, Phys. Rev. B
{\bf 76}, 035403 (2007).
 
\bibitem{PP03}
J.M. Pitarke and J.P. Perdew,  Phys. Rev. B {\bf 67}, 045101 (2003).
 
\bibitem{CPDG08}
L.A. Constantin, J.M. Pitarke, J.F. Dobson, A.                                        
Garcia-Lekue, and J.P. Perdew, Phys. Rev. Lett. {\bf 100}, 036401 (2008).


\bibitem{SSTP04}
V.N. Staroverov, G.E. Scuseria, J. Tao, and J.P. Perdew, Phys. Rev. B {\bf
69},  075102 (2004).
 
\bibitem{LT03}
B.J. Lynch and D.G. Truhlar, J. Phys. Chem. A {\bf 107}, 8996 (2003); {\bf
108}, 1460 (2004) (E).
 

\end{thebibliography}
\end{document}